THE EUROPEAN
PHYSICAL JOURNAL C

Regular Article - Experimental Physics

# Elliptic flow of identified hadrons in Au+Au collisions at $E_{lab}$ = 35 A GeV using the PHSD model

B. Towseef[1], M. Farooq[1], V. Bairathi[2,a] , B. Waseem[1], S. Kabana[2], S. Ahmad[1]

[1] University of Kashmir, Srinagar 190006, India
[2] Instituto de Alta Investigación, Universidad de Tarapacá, Casilla 7D, 1000000 Arica, Chile

**Abstract** We present predictions of elliptic flow ($v_2$) of identified hadrons at mid-rapidity ($|y| < 1.0$) in Au+Au collisions at $E_{lab}$ = 35 A GeV using the Parton Hadron String Dynamics (PHSD) model. The transverse momentum ($p_T$) dependence of identified hadron $v_2$ in minimum bias (0–80%) and three different centrality intervals (0–10%, 10–40%, and 40–80%) are presented. A clear centrality dependence of $v_2(p_T)$ is observed for particles and anti-particles. We also present the $p_T$ dependence of $v_2$ difference ($\Delta v_2$) between particles and corresponding anti-particles. A significant difference in $v_2$ values for baryons and anti-baryons is observed. Constituent quark scaling (NCQ) of $v_2$ is investigated in Au+Au collisions. We also present a $v_2(p_T)$ ratio between the HSD and PHSD modes to explore the effect of hadronic and partonic interactions in the medium. These predictions are useful for interpreting the data measured in the Beam Energy Scan (BES) program at RHIC. They will also be useful for the future Compressed Baryonic Matter (CBM) experiment at the Facility for Antiproton and Ion Research (FAIR) and Multi-Purpose Detector (MPD) at the Nuclotron-based Ion Collider facility (NICA).

## 1 Introduction

One of the primary goals of the relativistic heavy-ion experiments is the study of quark-gluon plasma (QGP), and the quantitative mapping of the QCD phase diagram [1,2]. The experiments at the Relativistic Heavy-Ion Collider (RHIC) [3,4], and the Large Hadron Collider (LHC) [5–7], have explored the QCD phase diagram in the region of high temperatures and vanishing baryon densities. RHIC has also performed a Beam Energy Scan (BES), including fixed target heavy-ion collisions that access high baryon density regions at small collision energies. The future CBM experiment at FAIR aims to investigate the region of the QCD phase diagram at high net baryon densities and moderate temperatures [8,9]. Also, the future MPD experiment at NICA plans to study collisions of heavy ions in the center-of-mass energy range between 4 and 11 GeV per nucleon to investigate the matter within high net-baryon densities [10,11].

The anisotropic flow of produced particles has long been considered a probe to study properties of the QCD matter created in heavy-ion collisions [12,13]. The anisotropic flow appears as a momentum-space anisotropy in the final states, and develops due to the pressure gradient resulting from the initial spatial anisotropy of the collision. Therefore, it is sensitive to the very early stages of the collision. Azimuthal anisotropy can be studied by the Fourier expansion of the azimuthal angle distribution of produced particles with respect to the reaction plane angle [14],

$$E\frac{d^3N}{dp^3} = \frac{1}{2\pi}\frac{d^2N}{p_T dp_T dy}\left(1 + 2\sum_{n=1}^{\infty} v_n \cos[n(\phi - \psi_{RP})]\right). \quad (1)$$

Here, $n$ denotes the harmonic order, $\phi$ is the particle's azimuthal angle, and $\psi_{RP}$ is the reaction plane angle made by the impact parameter vector and the beam direction. The $n^{th}$ order flow coefficient $v_n$ is given by the equation,

$$v_n = \langle\cos[n(\phi - \psi_{RP})]\rangle, \quad (2)$$

where $\langle\ \rangle$, represents an average over particles and events. The second order Fourier coefficient, also known as the elliptic flow ($v_2$). Due to the self-quenching nature of the initial spatial anisotropy that appears in the early stage of the collision, the elliptic flow $v_2$ remains conserved during the evolution of the system, therefore offers details on the dynamics at the beginning of the collision [15–17]. However, hadronic re-scattering at the later stages could alter the early dynamics of the collision. The identified hadron $v_2$ can be used to investigate the bulk properties of the medium produced in heavy-ion collisions. The elliptic flow has been studied previously in the following Refs. [4,5,18–23]. Multi-strange hadrons ($\Xi$ and $\Omega$) are expected to have smaller hadronic interaction cross-sections compared to non-strange hadrons [24]. Also, the kinetic freeze-out temperatures of multi-strange hadrons are close to the quark-hadron transition temperature suggested by the lattice QCD [23,25]. Therefore, it is believed that these multi-strange hadrons can predominantly provide information from the partonic stage of the collision [26–28].

[a] e-mail: vipul.bairathi@gmail.com (corresponding author)

In this paper, we report the first predictions on the elliptic flow of identified hadrons in Au+Au collisions at $E_{lab} = 35$ A GeV using the PHSD model [29,30]. The beam energy corresponds to the center of mass energy $\sqrt{s_{NN}} \approx 8.0$ GeV. The results are compared with the published elliptic flow measurements from the STAR experiment in Au+Au collisions at $\sqrt{s_{NN}} = 7.7$ GeV [31,32]. The elliptic flow is obtained as a function of transverse momentum in different centrality classes. The centrality dependence of the identified hadron elliptic flow is discussed. The constituent quark number scaling behavior of the elliptic flow is also studied. The effect of parton and hadron dynamics on the elliptic flow of identified hadrons will also be discussed.

The paper is organized as follows. The PHSD model and analysis method for calculating $v_2$ are briefly discussed in Sects. 2 and 3, respectively. The transverse momentum and centrality dependence of identified hadron $v_2$ are presented in Sect. 4. The difference in $v_2$ between particles and antiparticles is presented in Sect. 4.3. The number of constituent quark scaling of identified hadron $v_2$ is discussed in Sect. 4.4. Section 4.5 discusses the effect of hadronic and partonic interactions on $v_2$. Finally, we summarize the results reported in this paper in Sect. 5.

## 2 PHSD model

The PHSD model is a microscopic off-shell transport approach for strongly interacting systems in and out of equilibrium. PHSD incorporates both effective partonic and hadronic degrees of freedom and includes a dynamical description of the hadronization process from partonic to hadronic matter. It is based on a dynamical quasi-particle model (DQPM) [33,34] for partons developed to reproduce lattice quantum chromodynamics (lQCD) results, including the partonic equation of state, by introducing three DQPM parameters. These parameters are determined by comparing energy density, pressure, and entropy density from the DQPM to those from lattice QCD at $\mu_B = 0$ [35,36]. The hadronic phase is controlled via the Hadron String Dynamics (HSD) part of the transport method [37,38]. The PHSD approach completely realizes the evolution of a relativistic heavy-ion collision from the initial hard NN collisions out of equilibrium to the hadronization and final interactions of the resulting hadronic particles. PHSD has been effectively used to describe the data of p + p, p + A, and A + A reactions from SIS to LHC for energies ranging between 1 to 2760 GeV [39].

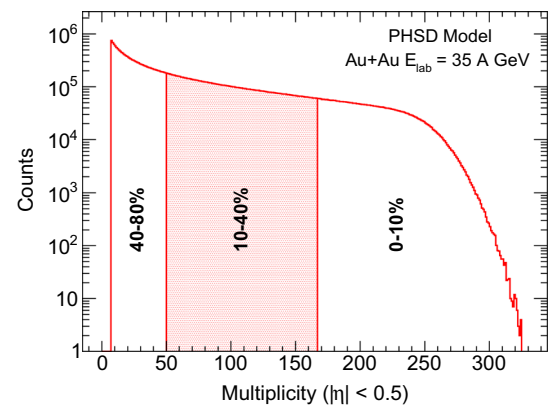

**Fig. 1** Reference multiplicity distribution ($|\eta| < 0.5$) in Au+Au collisions at $E_{lab} = 35$ A GeV from the PHSD model

For the results presented in this article, version 4.1 of the PHSD model is employed. The calculations are carried out with different modes of the model, namely the partonic (PHSD) and the hadronic (HSD) modes. We have simulated 50 million minimum bias Au+Au collision events at $E_{lab} = 35$ A GeV ($\sqrt{s_{NN}} \approx 8.0$ GeV). The data is generated for the impact parameter range, chosen randomly between $b_{min} = 0$ fm and $b_{max} = 15$ fm. The observables are calculated in the central rapidity region ($|y| < 1.0$) and different centrality intervals covering central to peripheral collisions. The centrality is calculated using the multiplicity within $|\eta| < 0.5$ from the PHSD model. The reference multiplicity distribution is shown in Fig. 1. The multiplicity is divided into three centrality classes 0–10%, 10–40%, and 40–80%.

## 3 Flow analysis method

The standard event plane method described in Ref. [14] is used for the elliptic flow analysis. The reaction plane angle $\psi_{RP}$ in Eq. 2 cannot be measured because the direction of the impact parameter is impossible to determine in the experiments. Therefore, an estimator of $\psi_{RP}$, namely the event plane angle $\psi_n$, is used to measure $v_2$. The estimated event plane angle is obtained from the azimuthal angle distribution of the produced particles. The $n$th harmonic event plane

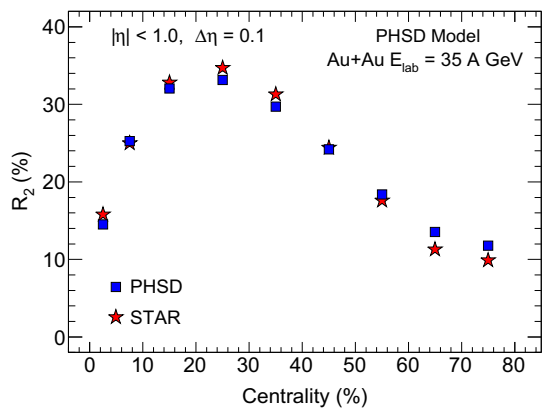

**Fig. 2** Event plane resolution as a function of centrality using the $\eta$-sub event plane method in Au+Au collisions at $E_{lab} = 35$ A GeV from the PHSD model

angle is defined as,

$$\psi_n = \frac{1}{n}\tan^{-1}\left[\frac{\sum_i \omega_i \sin(n\phi_i)}{\sum_i \omega_i \cos(n\phi_i)}\right]. \quad (3)$$

Here, $\phi_i$ is the azimuthal angle of $i$th particle, and $\omega_i$ is the weight taken as the $p_T$ of the particle to optimize the event plane resolution [14]. The sum runs over all the particles used to calculate the event plane angle in an event. In this work, $\psi_n$ is reconstructed using particles in the pseudo-rapidity range $|y| < 1.0$ and transverse momentum range $0.2 < p_T < 2.0$ GeV/c. The elliptic flow is obtained using the second-order event plane ($\psi_2$) and corrected for the event plane resolution ($R_2$). The $\eta$-sub event plane method is used with an $\eta$ gap of 0.1 between the two sub-events to suppress non-flow effects. The sub-events are defined in the negative ($-1 < \eta < -0.05$) and positive ($0.05 < \eta < 1$) pseudo-rapidity regions. $R_2$ is then estimated using,

$$R_2 = \sqrt{\langle\cos[2(\psi_2^a - \psi_2^b)]\rangle}, \quad (4)$$

where $\psi_2^a$ and $\psi_2^b$ are the sub-event plane angles in the negative and positive pseudo-rapidity regions. The elliptic flow for particles in the positive pseudo-rapidity window is calculated with respect to the $\psi_2$ calculated in the negative window and vice versa to minimize the auto-correlation effect. Additionally, we will present $v_2$ calculated with respect to the participant plane angle $\psi_2^{PP}$ provided by the PHSD model.

Figure 2 shows the event plane resolution as a function of centrality for Au+Au collisions at $E_{lab} = 35$ A GeV. The event plane resolution peaks near the mid-central collisions and decreases towards the central and peripheral collisions. It reaches a maximum of 34% for centrality (20–30%) in Au+Au collisions at $E_{lab} = 35$ A GeV. The decrease in resolution is attributed to the comparatively low multiplicity of peripheral collisions, whereas the effect is due to the small flow magnitudes for more central collisions. The resolution values from the PHSD model are very close to the published STAR results in Au+Au collisions at $\sqrt{s_{NN}} = 7.7$ GeV [31].

## 4 Results

In this section, we report $p_T$ and centrality dependence of the identified hadron $v_2$ at mid-rapidity ($|y| < 1.0$) in Au+Au collisions at $E_{lab} = 35$ A GeV ($\sqrt{s_{NN}} \approx 8.0$ GeV) using the PHSD model. We also compare $v_2$ results with the measurements from the STAR experiment in Au+Au collisions at $\sqrt{s_{NN}} = 7.7$ GeV. The results presented in this work are the first predictions of identified hadron $v_2$ from the PHSD model for future CBM experiment at FAIR and MPD experiment at NICA.

### 4.1 Differential elliptic flow $v_2(p_T)$

The elliptic flow $v_2$ as a function of $p_T$ for identified hadrons at mid-rapidity ($|y| < 1.0$) in minimum bias (0–80%) Au+Au collisions at $E_{lab} = 35$ A GeV from the PHSD model is shown in Figs. 3 and 4. The identified hadron $v_2$ increases with $p_T$ for all the studied particles. The obtained $v_2$ values are compared with the published results in Au+Au collisions at $\sqrt{s_{NN}} = 7.7$ GeV from the STAR experiment [31]. The $v_2$ of $\pi^\pm$, $K^\pm$, $K_s^0$, $p$, $\Lambda$, $\Xi^-$, and $\Omega^-$ are comparable to the STAR experiment data whereas $\bar{p}$ and $\bar{\Lambda}$ show larger deviations. The values of $v_2$ for $\bar{\Xi}^+$ from the STAR experiment are not significant enough for strong conclusions. The experimental $v_2$ values of $\bar{\Omega}^+$ is negative, therefore it is not shown in Fig. 4.

We also compare identified hadron $v_2$ calculated with respect to the event plane angle ($\psi_2$) and the participant plane angle ($\psi_2^{PP}$) in minimum bias Au+Au collisions at $E_{lab} = 35$ A GeV as shown in Figs. 3 and 4. The participant plane angle is provided by the PHSD model, which is calculated using the positions of participant nucleons. The $v_2$ values calculated with respect to $\psi_2^{PP}$ show larger deviation above $p_T > 1.0$ GeV/c. The deviation could be due to the event-by-event fluctuations in the positions of participating nucleons used to calculate $\psi_2^{PP}$. We also discuss $v_2$ calculated with respect to the event plane angle $\psi_2 = 0$ i.e. $v_2 = \langle\cos(2\phi)\rangle$. The $v_2$ values for mesons ($\pi$, $K$) above $p_T > 1.0$ GeV/c are lower when calculated using the $\psi_2 = 0$, while the magnitude of $v_2$ for baryons is similar within statistical uncertainties. The assumption of $\psi_2 = 0$ for each event does not account for the event-by-event fluctuations while calculating $\psi_2$, which affects meson $v_2$ at $p_T$ above 1.0 GeV/c in the PHSD model.

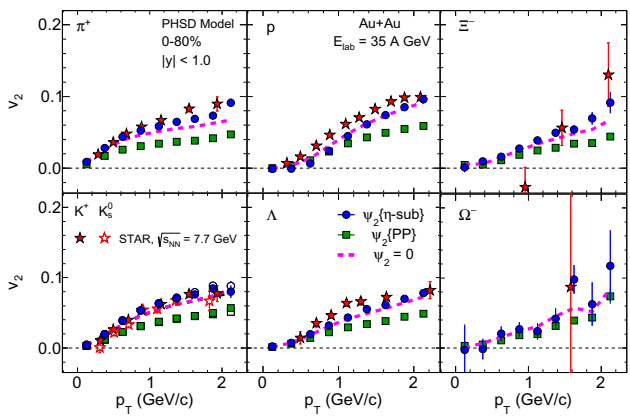

**Fig. 3** $v_2(p_T)$ for identified hadrons ($\pi^+$, $K^+$, $p$, $K^0_s$, $\Lambda$, $\Xi^-$, and $\Omega^-$) in 0–80% central Au+Au collisions at $E_{lab} = 35$ A GeV from the PHSD model (open and solid circles). The $v_2$ results from the STAR experiment in Au+Au collisions at $\sqrt{s_{NN}} = 7.7$ GeV (open and solid stars) are also shown [31]. The error bars shown are the statistical uncertainties

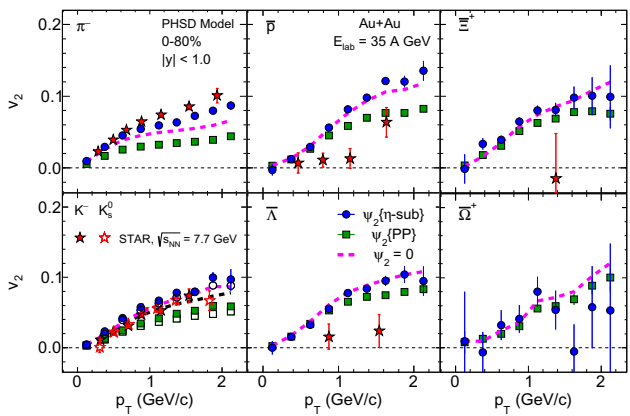

**Fig. 4** $v_2(p_T)$ for identified hadrons ($\pi^-$, $K^-$, $K^0_s$, $\bar{p}$, $\bar{\Lambda}$, $\bar{\Xi}^+$, and $\bar{\Omega}^+$) in 0–80% central Au+Au collisions at $E_{lab} = 35$ A GeV from the PHSD model (open and solid circles). The $v_2$ results from the STAR experiment in Au+Au collisions at $\sqrt{s_{NN}} = 7.7$ GeV (open and solid stars) are also shown [31]. The error bars shown are the statistical uncertainties

### 4.2 Centrality dependence of the elliptic flow

We report $v_2(p_T)$ for particles ($\pi^+$, $K^+$, $p$, $K^0_s$, $\Lambda$, and $\Xi^-$) and anti-particles ($\pi^-$, $K^-$, $\bar{p}$, $K^0_s$, $\bar{\Lambda}$, and $\bar{\Xi}^+$) at mid-rapidity ($|y| < 1.0$) for different centralities (0–10%, 10–40%, and 40–80%) in Au+Au collisions at $E_{lab} = 35$ A GeV from the PHSD model as shown in Figs. 5 and 6, respectively. A centrality dependence of $v_2$ is observed for all the hadron species. The magnitude of $v_2$ rises from central (0–10%) to peripheral (40–80%) collisions. A similar trend has been observed for $v_2$ of identified hadrons measured by the STAR experiment [32,40,41]. This centrality dependence is expected due to increased eccentricity of the initial overlap region of the colliding nuclei from central to peripheral collisions. The above observation is in line with the hydrodynamic

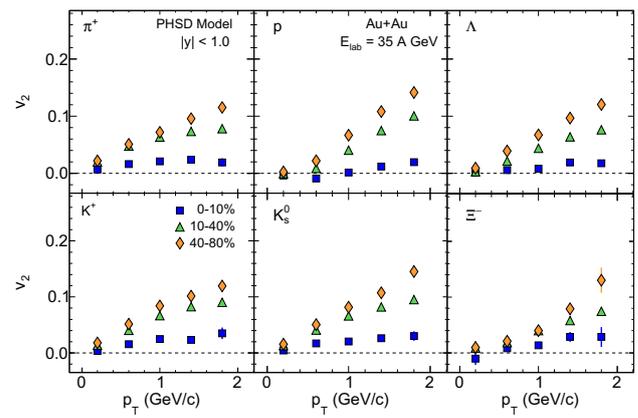

**Fig. 5** $v_2$ as a function of $p_T$ for identified particles ($\pi^+$, $K^+$, $p$, $K^0_s$, $\Lambda$, and $\Xi^-$) in 0–10%, 10–40%, and 40–80% centralities in Au+Au collisions at $E_{lab} = 35$ A GeV from the PHSD model. The error bars are statistical uncertainties

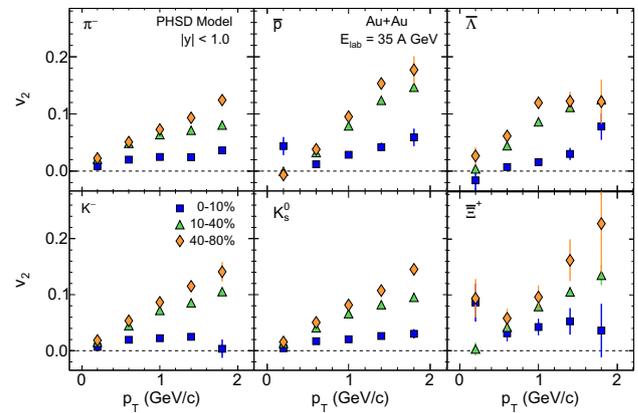

**Fig. 6** $v_2$ as a function of $p_T$ for identified anti-particles ($\pi^-$, $K^-$, $\bar{p}$, $K^0_s$, $\bar{\Lambda}$, and $\bar{\Xi}^+$) in 0–10%, 10–40%, and 40–80% centralities in Au+Au collisions at $E_{lab} = 35$ A GeV from the PHSD model. The error bars are statistical uncertainties

model, which predicts that the momentum anisotropy in the final state is driven by the initial spatial anisotropy [42].

### 4.3 $v_2(p_T)$ difference between particles and anti-particles

In Fig. 7, we present the $v_2$ difference ($\Delta v_2$) between the particles and their corresponding anti-particles. $\Delta v_2$ for pions and kaons do not exhibit any significant difference within statistical uncertainties in minimum bias Au+Au collisions at $E_{lab} = 35$ A GeV. This observation is expected for particles with the same mass and number of constituent quarks. There is a noticeable difference between $v_2(p_T)$ of baryons and anti-baryons, where $v_2(p_T)$ of anti-baryons is larger than that of the baryons in the PHSD model for minimum bias Au+Au collisions at $E_{lab} = 35$ A GeV. The difference between $v_2$ of baryons and anti-baryons increases with $p_T$. The difference could be attributed to the different baryons and anti-baryons production and absorption cross sections and to the momentum-dependent (anti-)baryons potential derived from the NL3 parametrization within the PHSD model framework [30,38,43–45]. The difference in $v_2$ of baryons could also be due to the baryon stopping at this lower beam energy. The same behavior is observed for $p$, $\Lambda$ and $\Xi$, which indicates that the exchange of one or more u-quark by s-quark does not have any impact on $\Delta v_2$.

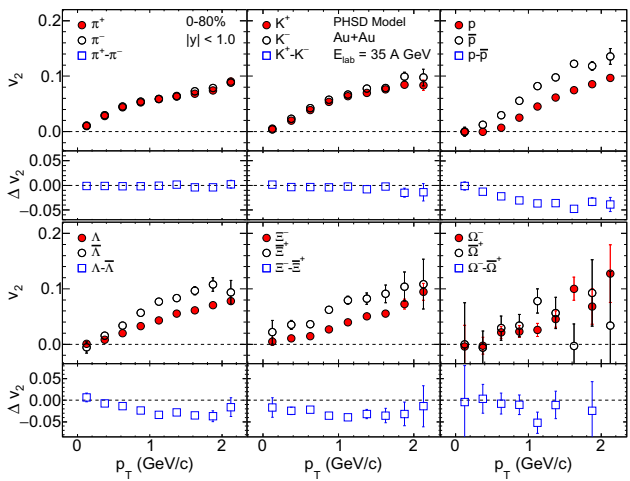

**Fig. 7** The elliptic flow $v_2$ as a function of $p_T$ for identified hadrons in 0–80% central Au+Au collisions at $E_{lab} = 35$ A GeV from the PHSD model. The difference in $v_2(p_T)$ between particles and anti-particles are displayed in the lower rows of each panel. The error bars are statistical uncertainties

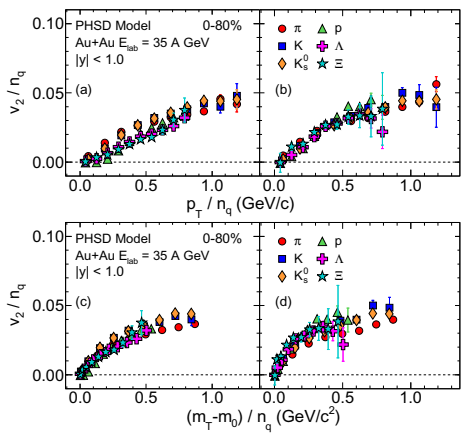

**Fig. 8** Number of quark scaled elliptic flow ($v_2/n_q$) as a function of $p_T/n_q$ and $(m_T - m_0)/n_q$ for identified particles (panel **a**, **c**) and anti-particles (panel **b**, **d**) in minimum bias Au+Au collisions at $E_{lab} = 35$ A GeV from the PHSD model. The error bars are statistical uncertainties

### 4.4 Number of constituent quark scaling

The elliptic flow $v_2(p_T)$ of identified hadrons at low $p_T$ follows the mass ordering predicted by the hydrodynamic model [46]. This mass ordering could be caused by the interplay of elliptic and radial flow, which reduces the $v_2$ magnitude for heavier particles at lower $p_T$. A baryon-meson splitting at intermediate $p_T$ is also shown in the experimental measurements [31,32]. The separation of $v_2$ for mesons and baryons at intermediate $p_T$ implies that $v_2$ is proportional to the number of constituent quarks ($n_q$). In order to investigate the number of constituent quark (NCQ) scaling of $v_2$ in the PHSD model, we have plotted $v_2/n_q$ versus $p_T/n_q$ and transverse kinetic energy $(m_T - m_0)/n_q$ as shown in Fig. 8 for minimum bias (0–80%) Au+Au collisions at $E_{lab} = 35$ A GeV. The transverse kinetic energy removes the dependence of the rest mass of particles on $v_2$ at lower $p_T$. The $v_2$ scaled with the number of constituent quarks for all the particles follows a single curve within the statistical uncertainties. The observed NCQ scaling of $v_2$ at this energy indicates that $v_2$ has been built up in the early stages of the collision where the dominating degrees of freedom are partonic in nature as suggested by the quark coalescence model for the QGP [47–49].

### 4.5 Hadronic and partonic interactions

Figure 9 shows the $p_T$ dependence of $v_2$ for identified particles and anti-particles in minimum bias Au+Au collisions at $E_{lab} = 35$ A GeV from the partonic and hadronic modes of the PHSD model. The partonic mode (also indicated as PHSD mode in the following) incorporates both the hadronic and partonic interactions, whereas the hadronic mode (HSD) includes only the hadronic interactions. The ratio of the $v_2$ values between the HSD and PHSD modes are shown in the lower rows of each panel. A value lower than one for the $v_2$ ratio indicates a sizeable effect of partonic interactions on the elliptic flow of hadrons within the PHSD model. We observed that the values of the ratio for $\pi^\pm$ and $K^\pm$ are $\sim$ 5–15% lower than unity, with a significance larger than $2\sigma$ for most $p_T$ points. The ratio values for anti-baryons ($\bar{p}$, $\bar{\Lambda}$, and $\bar{\Xi}^+$) are also lower than unity by $\sim$ 10–50%. The $v_2$ ratio for baryons ($p$, $\Lambda$, $\Xi^-$, and $\Omega^-$) is close to unity within the statistical uncertainties, which could be due to the baryon stopping process at lower beam energy such as $E_{lab} = 35$ A GeV.

## 5 Summary and conclusions

In this paper, we reported first predictions on the elliptic flow of identified hadrons ($\pi^+$, $K^+$, $p$, $K_s^0$, $\Lambda$, $\Xi^-$, and $\Omega^-$) and their anti-particles ($\pi^-$, $K^-$, $\bar{p}$, $K_s^0$, $\bar{\Lambda}$, $\bar{\Xi}^+$, and $\bar{\Omega}^+$) at mid-rapidity for minimum bias (0–80%) and different centrality intervals (0–10%, 10–40%, and 40–80%) in Au+Au collisions at $E_{lab} = 35$ A GeV from the PHSD model. The $v_2$ calculations are carried out using the $\eta$-sub event plane method. Non-flow correlations in the $v_2$ calculations were removed by introducing an $\eta$ gap between the positive and negative sub-events. The elliptic flow of the identified hadrons increases as a function of $p_T$. The $v_2$ increases from central to peripheral collisions for all the studied particle species. The observed centrality dependence of $v_2$ is consistent with the published results from the STAR experiment at RHIC. A significant difference in $v_2$ of baryons and anti-baryons is observed, increasing with $p_T$. The magnitude of $v_2$ for anti-baryons is larger compared to baryons at mid-rapidity for minimum bias Au+Au collisions in the PHSD model. The same behavior observed in the $\Delta v_2$ for $p(uud)$, $\Lambda(uds)$, and $\Xi^-(dss)$ shows that the exchange of one or more u-quarks by s-quarks is not responsible for this difference. The NCQ scaling is observed for all the particles and anti-particles in Au+Au collisions at $E_{lab} = 35$ A GeV in the PHSD model. The observed NCQ scaling of $v_2$ suggests that parton recombination might be responsible for the particle production and collectivity during the partonic stage of the medium created in Au+Au collisions at $E_{lab} = 35$ A GeV. This observation can be tested with the data from future heavy-ion collision experiments to determine the nature of the initially produced matter. These predictions are useful for understanding the data measured in the STAR BES program and also serve as the predictions for the future CBM experiment at FAIR and MPD experiment at NICA.

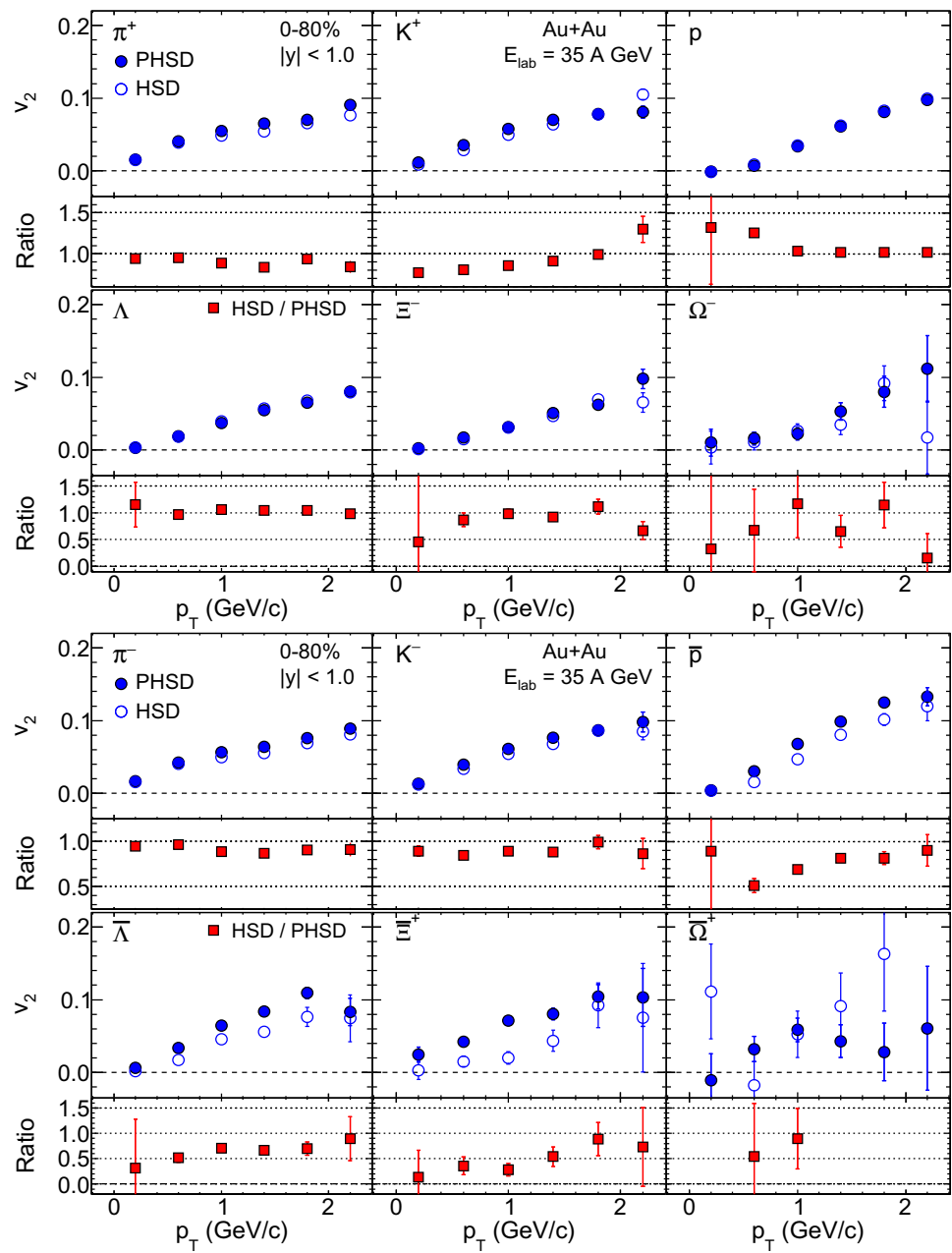

**Fig. 9** $v_2$ as a function of $p_T$ for identified particles ($\pi^+$, $K^+$, $p$, $\Lambda$, $\Xi^-$, and $\Omega^-$) and anti-particles ($\pi^-$, $K^-$, $\bar{p}$, $\bar{\Lambda}$, $\bar{\Xi}^+$, and $\bar{\Omega}^+$) in 0–80% central Au+Au collisions at $E_{lab} = 35$ A GeV from the partonic (PHSD) and hadronic (HSD) modes of the PHSD model. The ratio of $v_2(p_T)$ between HSD and PHSD is also shown in the lower rows of each panel. The error bars are statistical uncertainties

**Acknowledgements** S. Kabana acknowledges the financial support received from the Agencia Nacional de Investigación y Desarrollo

(ANID), Chile, from the ANID FONDECYT regular 1230987 Etapa 2023, Chile, and from the ANID PIA/ APOYO AFB220004, Chile. This research was partly supported by the cluster computing resource provided by the IT Division at the GSI Helmholtzzentrum für Schwerionenforschung, Darmstadt, Germany. The authors acknowledge helpful advice from the PHSD group members E. L. Bratkovskaya, V. Voronyuk, W. Cassing, P. Moreau, O. E. Soloveva, and L. Oliva.

**Data Availability Statement** This manuscript has no associated data or the data will not be deposited. [Authors' comment: This work is based on model simulation. The data for generating figures in this work can be provided upon request.]